\documentclass[manuscript,screen,acmsmall]{acmart}

\usepackage{CJKutf8}
\usepackage{array}
\usepackage{subcaption}  
\usepackage[most]{tcolorbox}
\usepackage{color}
\usepackage{soul}
\usepackage{multirow}
\usepackage{makecell}
\usepackage{tablefootnote}
\usepackage{multicol}
\usepackage{geometry}
\usepackage{listings}
\usepackage{xcolor}
\usepackage{tabularx}
\usepackage{caption}
\usepackage{boldline}
\usepackage{graphicx}
\usepackage{colortbl}
\usepackage{float}
\usepackage{booktabs}

\definecolor{red(ncs)}{rgb}{0.77, 0.01, 0.2}
\definecolor{tabcolor}{rgb}{.126,.126,.126}

\AtBeginDocument{%
  \providecommand\BibTeX{{%
    \normalfont B\kern-0.5em{\scshape i\kern-0.25em b}\kern-0.8em\TeX}}}

\setcopyright{acmlicensed}
\copyrightyear{2018}
\acmYear{2018}
\acmDOI{XXXXXXX.XXXXXXX}

\begin{document}

\title{Exploring the Danmaku Content Moderation on Video-Sharing Platforms: Existing Limitations, Challenges, and Design Opportunities}

\author{Siying HU}
\affiliation{%
  \department{Department of Computer Science}
  \institution{City University of Hong Kong}
  \city{Hong Kong SAR}
  \country{China}
}
\email{siyinghu-c@my.cityu.edu.hk}

\author{Zhicong Lu}
\affiliation{%
  \institution{City University of Hong Kong}
  \city{Hong Kong SAR}
  \country{China}
}
\email{zhiconlu@cityu.edu.hk}

\renewcommand{\shortauthors}{Hu, et al.}


\begin{abstract}
Video-sharing platforms (VSPs) have been increasingly embracing social features such as likes, comments, and Danmaku to boost user engagement. However, viewers may post inappropriate content through video commentary to gain attention or express themselves anonymously and even toxically. For example, on VSPs that support Danmaku, users may even intentionally create a ``flood'' of Danmaku with inappropriate content shown overlain on videos, disrupting the overall user experience. Despite of the prevalence of inappropriate Danmaku on these VSPs, there is a lack of understanding about the challenges and limitations of Danmaku content moderation on video-sharing platforms. To explore how users perceive the challenges and limitations of current Danmaku moderation methods on VSPs, we conducted probe-based interviews and co-design activities with 21 active end-users. Our findings reveal that the one-size-fits-all rules set by users or customizaibility moderation cannot accurately match the continuous Danmaku. Additionally, the moderation requirements of the Danmaku and the definition of offensive content must dynamically adjust to the video content. Non-intrusive methods should be used to maintain the coherence of the video browsing experience. Our findings inform the design of future Danmaku moderation tools on video-sharing platforms. 

\end{abstract}

%



\received{20 February 2007}
\received[revised]{12 March 2009}
\received[accepted]{5 June 2009}

\maketitle

\section{Introduction}

Video-sharing platforms (VSPs) have been increasingly embracing social features such as liking and commenting to boost participant engagement. The rise of Danmaku (i.e., a form of interactive commentary overlaid directly onto videos that resembles a ``bullet curtain" or ``barrage''), however, marked a significant shift in how digital content was consumed and interacted with on VSPs \cite{wu2018danmaku}. 
Danmaku's defining characteristic is its time-synchronization  \cite{wu2018danmaku}, wherein participants' reactions are directly linked to specific scenes or events within a video. 
This synchronization creates a layered viewing experience where the original video content and participant-generated comments coexist, enhancing the depth of engagement.

\begin{figure}
    \centering
    \includegraphics[width=1\textwidth]{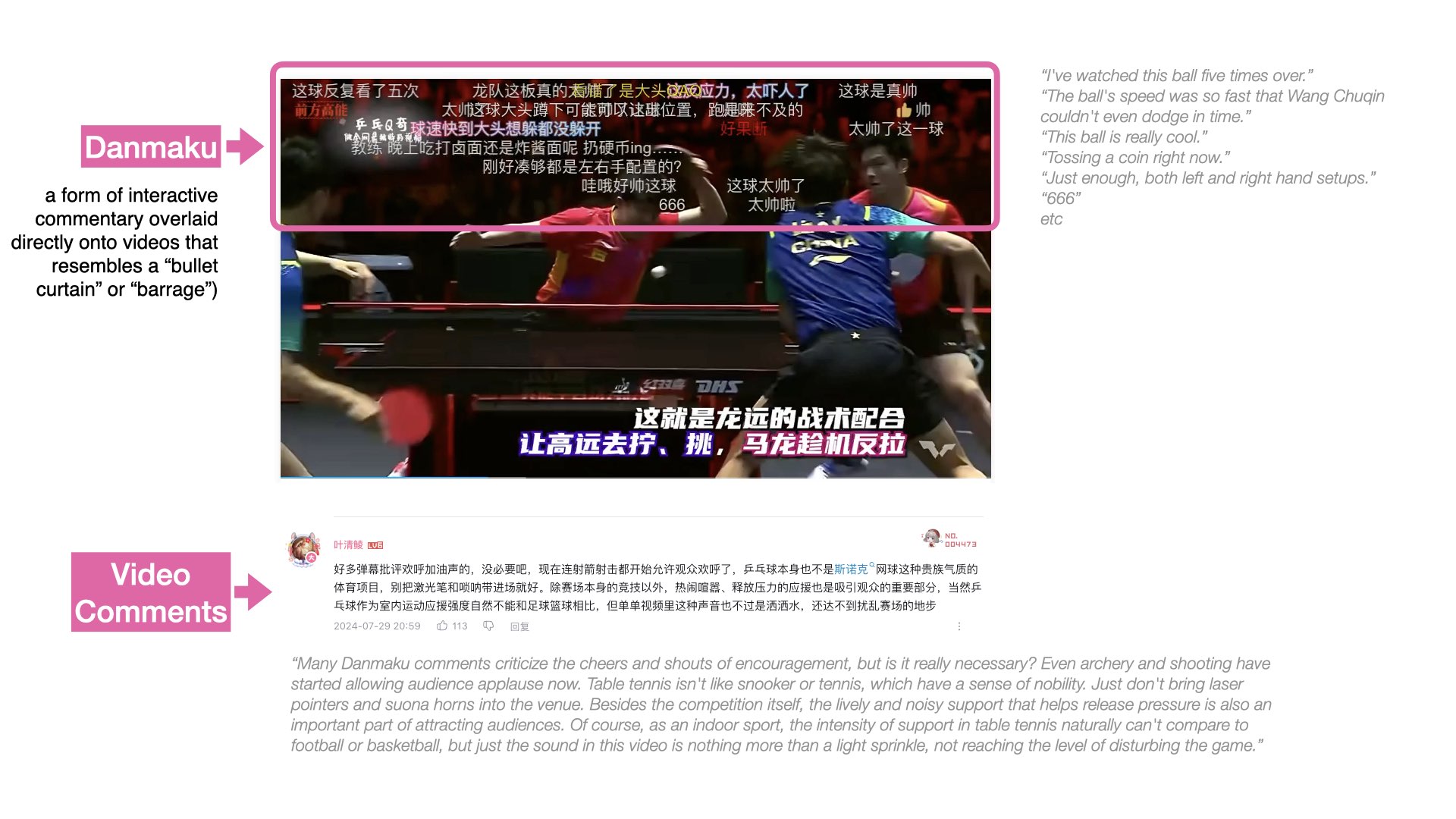}
    \caption{Two video commentary types in Video Sharing Platforms (VSPs). Bottom: Conventional comments appear in a dedicated section below videos, separating viewing and commenting into asynchronous activities. Upper: Danmaku comments overlay the video content and move across the screen for several seconds, allowing viewers to engage with comments synchronously while watching.}
    \Description{A comparison of two video commentary systems: traditional comments shown at the bottom of the video page, and Danmaku comments overlaid directly on the video content. The Danmaku comments float across the screen, enabling real-time viewer interaction during video playback.}
    \label{fig:example_VSPsDanmaku}
\end{figure}

Danmaku's synchronized, overlain presentation fosters a sense of shared viewing, as participants can see and respond to each other's reactions in real-time as the video unfolds. It also creates a vibrant, albeit sometimes chaotic, communal atmosphere, distinguishing it from the more structured and less intrusive comment systems found on platforms like YouTube or Twitch (see \autoref{fig:example_VSPsDanmaku} for an example ).

Existing studies have highlighted how Danmaku can transform the solitary act of video watching into a socially enriched experience, fostering a sense of community among viewers, such as co-viewing \cite{fang2018co} or companion-style learning \cite{wang2021study,chen2021learning}, and foster more conversation in non-gaming contexts \cite{lu2021understanding}. Lin et al. explored its use in online lectures, finding that Danmaku enhanced learners' interactions and course engagement \cite{lin2018exploratory}. Danmaku has also been shown to promote participant participation and facilitate community-building through the use of linguistic memes and anonymous, positive comments \cite{wu2019danmaku}.
Furthermore, research by Ma and Cao explored the impact of Danmaku on participant engagement, noting that the immediacy and visibility of comments often enhanced the emotional and entertainment value of video content \cite{ma2017video}. While Danmaku has enabled viewers to express themselves in new ways, viewers may unintentionally create a ``flood'' of Danmaku with inappropriate content, which can harm others and disrupt the overall viewer experience \cite{lapidot2012effects}. There thus, is a need to moderate Danmaku content on VSPs.

Danmaku systems often employ a hybrid moderation approach that combines automated filters with human oversight to manage participant-generated comments in real time. The moderation process typically involves both platform staff and volunteer moderators, who monitor content for rule compliance, remove inappropriate comments, and interact with the community to maintain a positive environment. These volunteers follow a set of guidelines provided by the platform and may undergo training to familiarize themselves with moderation tools and procedures. They often provide general community oversight but can also assist during live events or appeals handling. 

Most prior work on VSP content moderation has adopted a technical, platform-centric view, overlooking critical aspects such as how viewers experience and perceive current Danmaku moderation approaches, the nuanced ways community dynamics and contextual factors influence definitions of acceptable and unacceptable content, and viewer needs and requirements for more inclusive, context-aware, and seamless moderation experiences. Motivated by these gaps, three research questions guided this research:

\begin{itemize}
     \item \textbf{RQ1:} What are the perceived limitations of existing Danmaku moderation approaches from viewers' perspectives?
     \item \textbf{RQ2:} What challenges do viewers encounter when they moderate or are exposed to inappropriate Danmaku on VSPs?
     \item \textbf{RQ3:} What design opportunities do participants identify to enhance Danmaku moderation methods?

\end{itemize}

RQ1 aims to explore the constraints inherent in current systems' functionalities, designs, or implementations. These limitations may stem from technological restrictions, platform policies, or system design philosophies. In contrast, RQ2 focuses on the difficulties and obstacles viewers face during their practical interactions. These challenges may arise from viewer-system interactions or personal and social issues when dealing with inappropriate content, reflecting real-world user experiences. While these questions may appear similar, they offer complementary insights from system and user perspectives respectively. By differentiating between system limitations and user challenges, we aim to gain a more holistic understanding of the complexities in Danmaku moderation and propose more comprehensive and effective improvements in subsequent research.

Our work does not aim to develop a complete new moderation system/algorithm. Rather, the goal is to gain insights from users' perspectives to identify gaps in current approaches and explore design opportunities for more effective, user-friendly, context-aware Danmaku moderation tools and strategies. While prior work has advanced moderation methods \cite{kolla2024llm,cai2022understanding,ghosh2024aegis,hu2024danmodcap}, there is a lack of understanding of how users actually experience and perceive these methods in the Danmaku context. A user-centered, participatory design study is needed to uncover real-world challenges, cultural nuances, and innovative opportunities that purely technical approaches may overlook.
By employing probe-based interviews and co-design activities with 21 Internet-savvy users, we seek to gain a nuanced understanding of their perspectives on the challenges and limitations of current Danmaku moderation methods. The use of probe-based interviews allows us to delve deeply into users' genuine interactions with Danmaku, capturing their real-time reactions and nuanced feedback on moderation practices. This method enables us to uncover subtle user needs and pain points that conventional survey methods might miss. 
Furthermore, the co-design activities involve users in the creative process of developing new moderation tools and strategies. This participatory approach not only empowers users by valuing their input but also ensures that the resulting solutions are closely aligned with their actual needs and expectations. By collaborating directly with users, we could identify innovative moderation opportunities that are not only technically feasible but also practically effective in enhancing the user experience.

Through these studies, we strive to understand better the unique challenges in moderating real-time, contextualized user-generated video overlays like Danmaku. Our findings can inform the design of more effective, proactive, and visionary moderation tools that balance open expression with eliciting online pro-social behaviours for all viewers.

\section{Background and Related Work}
In this section, we provide an overview of the existing literature pertaining to the live Content moderation in online communities. Furthermore, we reviewed the user-centered design methodologies tailored to content moderation.

\subsection{Live Content Moderation in Online Communities}
Live content moderation is a crucial aspect of maintaining a safe and positive environment in online communities. It involves the real-time review and removal of harmful or inappropriate content, such as hate speech, bullying, and misinformation \cite{cai2021moderation,scheuerman2021framework}. Several research papers highlight the importance and challenges of live content moderation. The dynamic and interactive nature of live streaming platforms necessitates immediate attention to user-generated content, which can be both voluminous and rapidly changing. Cai et al., \cite {cai2021moderation} explored the voluntary moderation practices within live streaming communities, emphasizing the cognitive overload and emotional toll faced by moderators who must make swift decisions with limited guidance. The study identifies a range of strategies employed by moderators, from proactive measures like declaring presence and rule echoing to reactive strategies involving viewer profiling and conflict management within moderation teams. Seering et al.,'s work examines the role of research in supporting community-based models for online content moderation, advocating for a more collaborative approach between platform designers and community members \cite{seering2020reconsidering}. This includes the development of chatbots as community members that can assist in moderation tasks \cite{seering2019beyond}. Ghosh et al., propose AEGIS, a novel online adaptation framework that leverages these LLM experts to perform dynamic content moderation with strong theoretical guarantees, adapting to changing data distributions and policies in real-time \cite{ghosh2024aegis}. Kolla et al,. investigates the feasibility of employing LLMs to identify rule violations on Reddit \cite{kolla2024llm}. The researchers examine an LLM-based moderator's reasoning capabilities across diverse subreddits and posts, revealing that while the system has a high true-negative rate, its true-positive rate is significantly lower, indicating a tendency to flag keyword-matching violations but struggle with more complex cases. This research underscores the need for careful integration of LLMs into content moderation workflows and highlights the importance of designing platforms that effectively support both AI-driven and human-in-the-loop moderation strategies.

\begin{figure}
    \centering
    \includegraphics[width=1\textwidth]{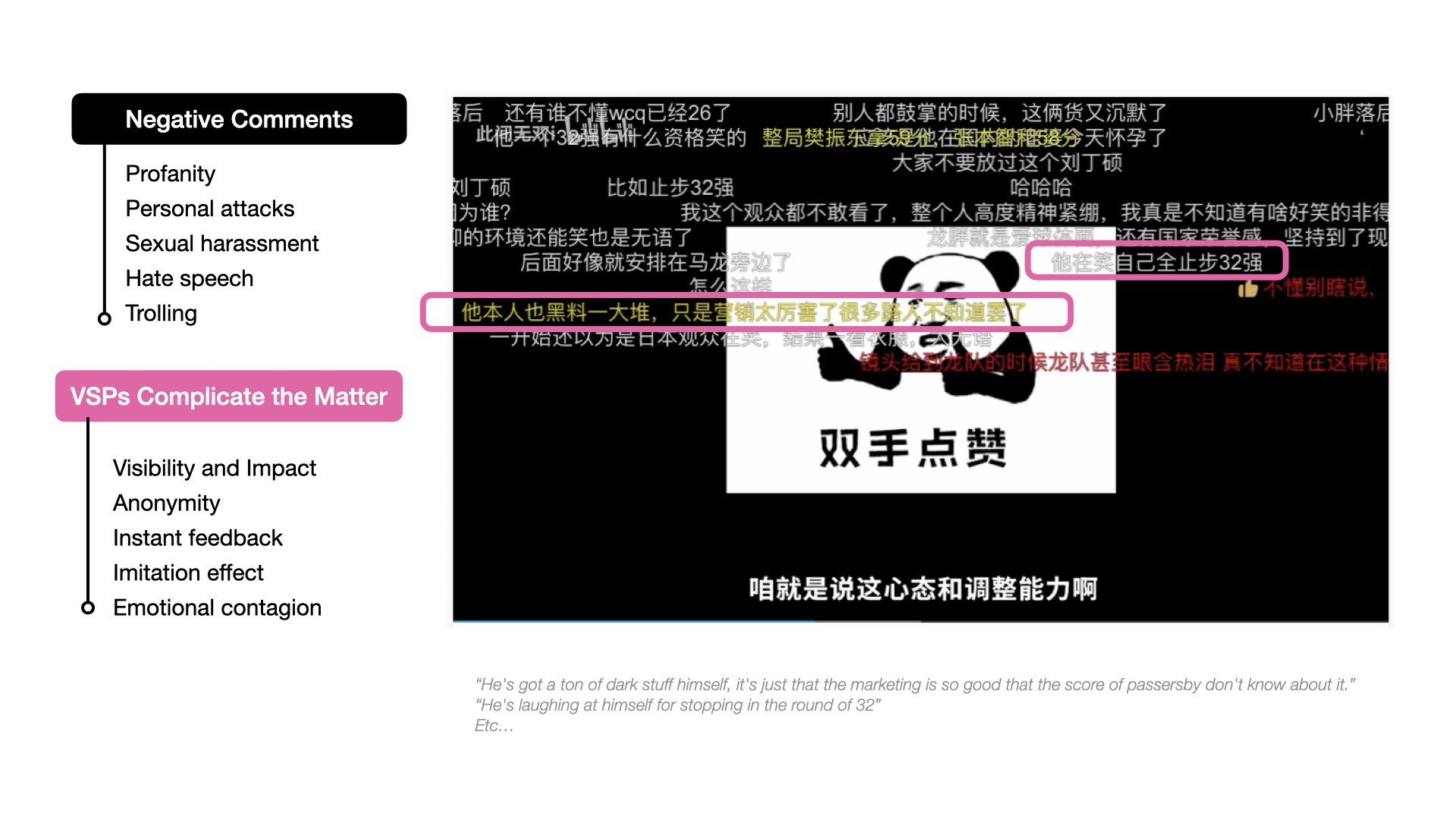}
    \caption{Example of Danmaku Anti-Social Behaviours and Potential Negative Impacts in VSPs.}
    \Description{The figure shows examples of anti-social behaviors in Danmaku comments on video sharing platforms, illustrating how these negative interactions can impact the viewing experience and community atmosphere.}
    \label{fig:example_AntiDanmaku}
\end{figure}

However, balancing automated moderation approaches with human judgment is key, as automated systems are efficient but lack context, while manual moderation is resource-intensive \cite{chang2014aligning}. Despite their proficiency in handling large volumes of content, occasionally falter in comprehending the context, thereby risking the inadvertent promotion of inappropriate conduct \cite{lai2022human}. Such limitations have ignited a surge of interest in the integration of human expertise with AI capabilities within the content moderation field. This collaborative model seeks to capitalize on the respective advantages of human judgment and AI efficiency, forging a more potent and nuanced approach to moderating online content \cite{jhaver2019human}. User agency in content moderation is gaining traction, with personal moderation tools empowering users but potentially leading to echo chambers. Platform moderation enforces standardized rules, but it can be perceived as overreaching \cite{seering2019moderator}. User agency in content moderation plays a critical role in this balance, offering personalization while posing challenges in ensuring diverse and healthy discourse \cite{chang2014aligning}. Understanding user motivations, trust, and the dynamics of participation is essential in designing effective moderation strategies that cater to the needs of diverse online communities \cite{cusumano2020future,chang2014aligning}.

Content moderation in real-time, especially in the context of Danmaku, poses significant challenges due to its rapid, dynamic, and often unpredictable nature (examples can see in \autoref{fig:example_AntiDanmaku}). The traditional dichotomy of automated versus manual moderation systems, as discussed by \cite{jhaver2023personalizing}, reveals the inadequacy of existing approaches in handling the nuanced and context-sensitive nature of Danmaku. While automated systems are capable of filtering explicit content quickly, they often lack the subtlety to understand cultural nuances and sarcasm, leading to over-moderating or misinterpretation. The use of BERT architecture for classifying Danmaku comments into video categories presents a novel approach to managing irrelevant comments and potentially improving content quality \cite{tamura2021selective}. On the other hand, manual moderation, despite its higher accuracy in context comprehension, struggles with scalability and real-time response, as mentioned in the work of \cite{he2021beyond,lu2018streamwiki}. This complexity necessitates a hybrid approach that balances efficiency with contextual sensitivity. The challenge of moderating real-time Danmaku content in video-sharing platforms requires a nuanced approach that can adapt to the platform's unique interactive dynamics, which mentioned by \cite{niu2023building}. Automated systems need to be complemented with manual oversight to ensure context-sensitive and culturally aware moderation, which is critical in preserving the intended social interaction and community engagement that Danmaku facilitates.
Building on the existing work mentioned above, the specific gaps or limitations in how prior work has handled the unique challenges posed by Danmaku's real-time, context-dependent, and community-driven nature is lack of explore. Our research focuses on the content moderation challenges of real-time updating and social functionality associated with Danmaku comments, as well as an exploration of the limitations of current moderation methods.

\subsection{Participation-Driven Design to Envision Content Moderation}

HCI researchers have widely utilized participatory and co-design methods to craft digital technology with the end-users' perspective in mind \cite{harrington2019deconstructing,pierre2021getting,zhang2022algorithmic}. These approaches excel in uncovering the challenges that users face and in providing solutions that are specifically tailored to meet their needs. They significantly reduce the barriers to user adoption. By including users in the design process, researchers are able to incorporate their specialized insights, thereby enhancing the user-friendliness of the technology

The co-design approach in formulating content moderation strategies introduces a collaborative framework that incorporates inputs from diverse stakeholders including platform developers, content creators, and the audience \cite{guo2017user,agha2023strike,storms2022transparency}. Some studies involving end-users in the design process of moderation tools can lead to more nuanced and user-friendly solutions, aims for transparency \cite{storms2022transparency} , personalizing \cite{jhaver2023personalizing}, risk intervention \cite{agha2023strike} and etc. This approach is particularly effective in addressing the unique challenges posed by Danmaku, as it allows for a more granular understanding of user behavior and community standards. The research exemplifies this by demonstrating how user feedback helped refine algorithmic moderation tools to better accommodate the fast-paced and varied nature of Danmaku, striking a balance between maintaining user engagement and ensuring a non-toxic community environment \cite{guo2017user}. Riedl, Christoph, et al. exemplifies how user feedback helped refine algorithmic moderation tools to better accommodate the fast-paced and varied nature of Danmaku \cite{riedl2013effect}. Their study on designing rating scales for evaluating user-generated content in online innovation communities reveals the importance of user perceptions and decision quality, which is crucial for maintaining user engagement and ensuring a non-toxic community environment.
The use of co-design not only addresses the immediate issues of content moderation but also fosters a sense of ownership and responsibility among the users. This aspect is crucial in cultivating a self-regulating community, as highlighted by \cite{gebauer2013dark} in their study on participatory design in online spaces. Their research underscores the long-term benefits of engaging users in policy formation, leading to more sustainable and adaptable moderation practices. The co-design approach in content moderation, particularly for Danmaku in online communities, leads to more effective, user-centric solutions by involving users in the design and refinement of moderation tools. This approach not only enhances the immediate moderation process but also contributes to the long-term sustainability and adaptability of community governance.

while prior technical work has advanced moderation methods, there is a lack of understanding of how users actually experience and perceive these methods in the Danmaku context. A user-centered, participatory design study is needed to uncover real-world challenges, cultural nuances, and innovative opportunities that purely technical approaches may overlook. Drawing on previous collaborative design practices \cite{kim2024unlocking}, we utilized this approach in our research to explore the design support for content moderation of Danmaku comments on video streaming platforms, in order to contribute design guidelines for this purpose.

\section{Study 1: Probe-based Interview}
\label{ProbInterview}

To understand the existing challenges (RQ1) and perceived limitations (RQ2) with Danmaku moderation, we conducted probe-based interviews with active viewers on VSPs. 

\subsection{Participants and Recruitment}

This study recruited participants with and without experience with Danmaku, a real-time commenting system that overlays comments on video content. Danmaku is a built-in feature in many Video Sharing Platforms (VSPs) such as TikTok, Xiaohongshu, Bilibili, and WeTV, often enabled by default to increase user engagement. We recruited participants through social media, local school mailing lists, and other channels. Potential participants were invited to complete a screening questionnaire about their previous experience with VSPs and Danmaku-related features, as well as their knowledge of content moderation. We recruited a total of 21 participants (\autoref{tab:participant}) with backgrounds in computing, design, journalism and communication, chemical biology, psychology, and law. They were all non-native English speakers, but their English proficiency exceeded the IELTS 6.5 standard. To ensure comprehension and reduce language barriers, they were provided with video content in their mother languages, which allowed them to engage with the material more naturally and effectively. When dividing participants into groups for collaborative experiments, we carefully balanced the distribution of experienced and inexperienced Danmaku users. This approach ensured a diverse range of perspectives within each group. The experiment was conducted in an offline lab at a local university.

While VSP viewers may not actively moderate content, their firsthand experiences make them well-positioned to identify moderation issues and contextual nuances that platform- or moderation-centric approaches may miss (e.g., cultural nuances, community dynamics). They can articulate the implicit needs and desires that shape their perceptions of what constitutes acceptable versus unacceptable content. Moreover, they can provide insights into how moderation approaches impact their overall engagement and viewing experience. By including both experienced and inexperienced Danmaku users, our study aims to capture a comprehensive understanding of content moderation challenges and potential solutions in VSPs, particularly in the context of real-time commenting systems like Danmaku.

\subsection{Technology Probes Preparing}
To create the technology probes for study, we first systematically observed the most popular Danmaku moderation features and interactions on multilingual VSPs. The observed platforms were selected based on the Similarweb global rankings of dedicated video-hosting websites\footnote{We focused on popular VSPs that supported Danmaku as of July 23, 2023 
(https://en.wikipedia.org/wiki/List\_of\_online\_video\_platforms)}, including YouTube, TikTok, Bilibili, Twitch, Niconico, and so on.  We included platforms that could offer a variety of video formats and had consistent moderation features when selecting the VSPs to use.

Drawing on previous work investigating content moderation tools \cite{jhaver2023personalizing}, we created a new account or used an existing account on each platform and viewed its video playing interface and settings page regarding Danmaku moderation to observe the options available to viewers. We also reviewed the options that were available through third-party moderation tools such as acghelper\footnote{Acghelper: https://acghelper.com/}, TwitchChatDanmaku\footnote{TwitchChatDanmaku: https://github.com/wheatup/TwitchChatDanmaku?tab=readme-ov-file} and YouTube LiveChat\footnote{YouTube LiveChat: https://chromewebstore.google.com/detail/youtube-livechat-flusher}. Given our focus on moderation tools for Danmaku, We did not consider playback-based settings such as Danmaku visual modifications (e.g., font size or color) and video layout (e.g., occupying 1/3 of the screen or displaying it only at the top). 

We observed and analyzed different video types, each with unique characteristics that cater to different viewer preferences and Danmaku moderation habits. \textit{Horizontal videos} typically featured a landscape orientation, which is the traditional format for movies, TV shows, and online videos that are designed to be viewed on larger screens or in a horizontal orientation platforms (e.g., YouTube, Bilibili, and WeTV). This format was well-suited for storytelling that benefits from a wider field of view. \textit{Vertical videos}, in contrast, were oriented in portrait mode, which has gained popularity on social media platforms (e.g., Tiktok) where viewers often hold their phones upright. This format was ideal for intimate, personal content and easily consumable in a vertical scrolling environment. \textit{Long video} content was characterized by its extended duration, allowing for comprehensive storytelling and an in-depth exploration of subjects. These videos were commonly found on traditional media outlets and online streaming platforms and provided a rich and immersive viewing experience. Conversely, \textit{short videos} were distinguished by their brevity, making them ideal for platforms that prioritize quick, engaging content that can be easily consumed on mobile devices in a fast-paced, on-the-go context. Additionally, we considered \textit{livestreaming}, a dynamic format that allowed content to be broadcast in real-time, fostering a sense of immediacy and interactivity. Livestreaming was particularly popular for events, conversations, and demonstrations where the live element was valuable for audience engagement and participation.

\begin{figure}
    \centering
    \Description{A figure showing four different types of moderation probes. Panel A illustrates Personal Control Moderation, Panel B shows Reactive Moderation, Panel C demonstrates Community-Driven Moderation, and Panel D presents Distributed Moderation.}
    \includegraphics[width=1\textwidth]{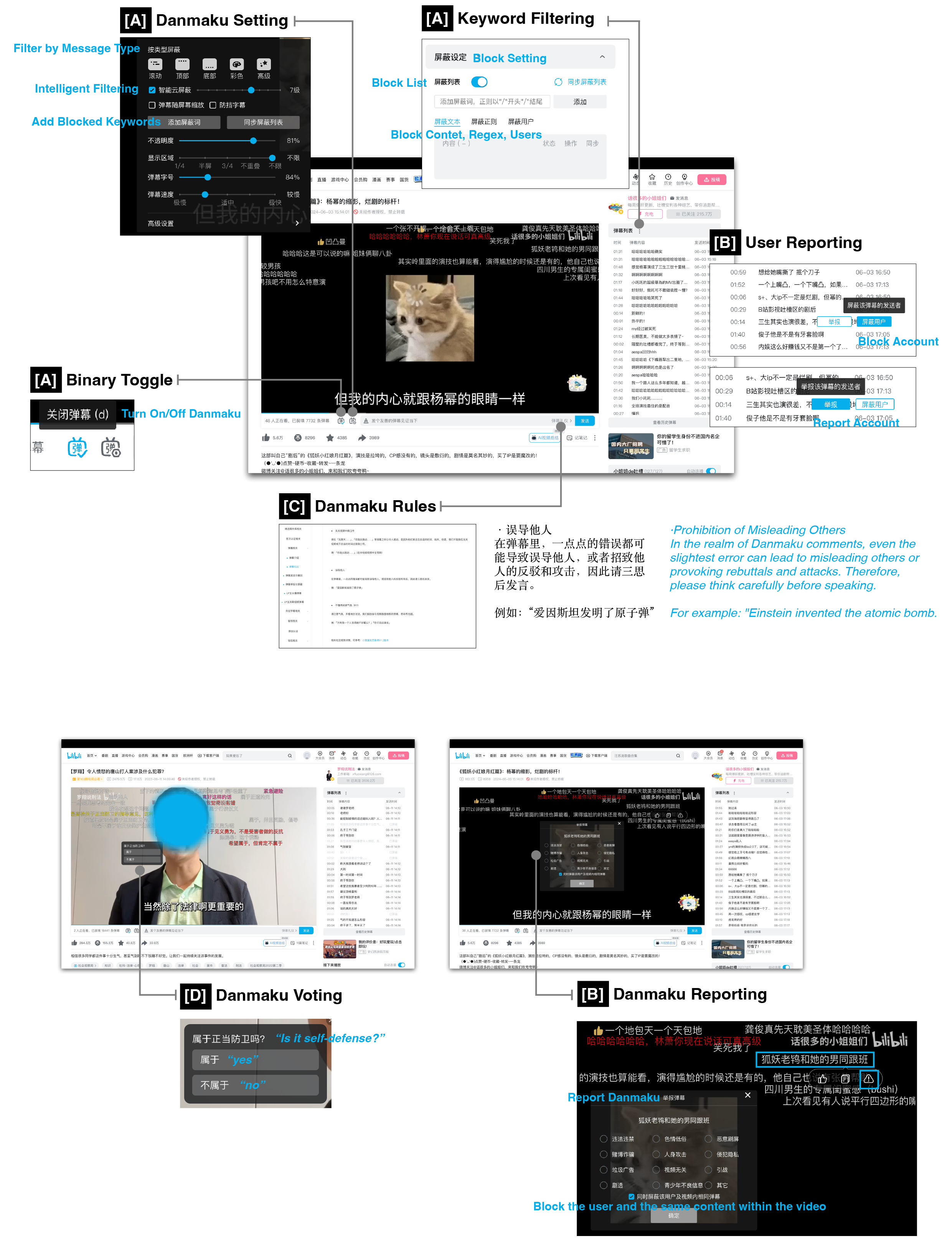}
    \caption{Example of the collected probes, which included [A] Personal Control Moderation, [B] Reactive Moderation, [C] Community-Driven Moderation, and [D] Distributed Moderation.}
    \label{fig:probes_example}
\end{figure}

Through these observations, we summarized and classified the most four common Danmaku moderation methods (i.e., Personal Control, Reactive, Community-Driven, and Distributed) based on these different types of videos (\autoref{fig:probes_example}):

\begin{itemize}
    \item \textbf{Personal Control Moderation}: This approach empowered individual participants to control the content they were exposed to, providing a personalized moderation experience. Such moderation included \textit{Intelligent Filtering}\footnote{Intelligent Filtering: An automated feature that used preset rules and algorithms to identify and filter inappropriate Danmaku without human intervention.}, \textit{Keyword Filtering}\footnote{Keyword Filtering: customized keywords can be used to create display rules for Danmaku, such as blocking comments from specific participants or containing certain words.}, and \textit{Binary Toggle}\footnote{Binary Toggle: enabling or disabling Danmaku comment filtering with a simple on/off switch, allowing for quick adjustments based on one's current needs.}
    
    \item \textbf{Reactive Moderation}: This strategy involved taking action after inappropriate content had surfaced, relying on the proactive identification and handling by viewers or moderators, such as \textit{Viewer Reporting\footnote{Viewer Reporting: viewers could select and report Danmaku for review that they considered inappropriate or offensive, contributing to community-based content oversight.}}.
    
    \item \textbf{Community-Driven Moderation}: Focused on self-governance within the community, this method used rules established by community members to maintain order, e.g., \textit{Danmaku Etiquette\footnote{Danmaku Etiquette: viewers self-regulate their behavior by adhering to guidelines that promoted a respectful and constructive commenting environment.}}.
    
    \item \textbf{Distributed Moderation}: This approach decentralized the moderation task among community members, leveraging a collective effort to manage content, e.g., \textit{Danmaku Voting\footnote{Danmaku Voting: A community-driven process where participants participated in moderation through voting, allowing for democratic decision-making on content acceptability.}}.
\end{itemize}

We then selected the videos that each participant would watch. The genres covered Game or Esports content, Tutorial videos, Vlogs or Lifestyle, Anime, and Short videos. These videos represented the most popular genres Bilibili and TikTok, as these platforms encompassed the most common Danmaku moderation approaches we analyzed and featured similar video types. The videos representing each genre were selected by ourselves and a fellow researcher based on similar levels of Danmaku counts, video duration, and Danmaku moderation features. Each genre of video material could be found on both platforms. Bilibili was used as it is the largest video website in China with Danmaku and had a rich source of videos for the study. TikTok, a global short video platform, had a wide participant base that spanned different cultures and regions, offering a diversified perspective for our research. 

\begin{table}
    \centering
    \resizebox{\textwidth}{!}{
    \begin{tabular}{ccccc}
    \toprule[1.2pt]
         \textbf{Group}&  \textbf{Participant}&  \textbf{Gender/Age} &\textbf{\makecell[c]{Video Sharing Platforms \\Frequently Used}} &\textbf{\makecell[c]{Self-identified \\Danmaku Experience}}\\
    \toprule[1.2pt]
         G1&  P1&  Felmale/29& BILI, XHS, TT & High\\
         &  P2&  Felmale/27& BILI, IQ, TV & High\\
         &  P3&  Male/25& XHS, TT, HY & Medium\\ \hline
         G2&  P4&  Male/27& YT, XHS, TW & Low\\
         &  P5&  Felmale/29& BILI, XHS, TV & High\\
         & P6& Male/24& BILI, TW, YT &High\\ \hline
         G3& P7& Male/25& BILI, XHS, AF &High\\
         & P8& Felmale/29& BILI, IQ, TV & Medium\\
         & P9& Male/27& XHS, TT, TV & Medium\\ \hline
         G4& P10& Male/25& BILI, HY, TT & Low\\
         & P11& Male/29& XHS, TT, BILI & Medium\\
         & P12& Felmale/22& BILI, TT, IQ & High\\ \hline
         G5& P13& Felmale/25& BILI, XHS, TT & High\\
         & P14& Felmale/25&  BILI, XHS, TT & Low\\
         & P15& Felmale/29& XHS, TT, TV & High \\ \hline
         G6& P16& Male/23& TW, BILI, YT & Medium \\
         & P17& Male/28& BILI, XHS, TV & High\\
         & P18& Male/31& TW, YT & Low  \\ \hline
         G7& P19& Felmale/25& BILI, XHS, TT & High \\
         & P20& Male/31& BILI,  & Low\\
         & P21& Felmale/26& TT, HY, TV & Medium \\ \hline
    \end{tabular}
    }
    \caption{Participant's Demographic Information. BILI = Bilibli, XHS = Xioahongshu, TT =Tiktok, TW =Twitch, YT = YouTube Live, HY = Huya, IQ= iQIYI, TV= Tencent Video, AF = AcFun;}
    \label{tab:participant}
    
\end{table}

\subsection{Procedure}
The interviews lasted around 90 minutes, with an audio recorder and screen recording tool being used record the interviews. At the beginning of each interview, each participant was asked about their own behavior when using VSPs, their preferences for different types of videos, and their browsing practices. Next, they were asked about their experiences related to Danmaku, how they generally manage Danmaku, and the causes and the effects of control when using VSPs. They were then asked if they had encountered offensive Danmaku, how they had responded to the content, what actions they had taken in such cases, if any, and what the outcome had been. After completing these questions, we briefly reviewed the results with the participants and asked them if they had anything to add to prevent omissions or misunderstandings during the initial formulation.

Next, we briefly reviewed the goals of the experiment and asked participants to open the different probes to familiarize themselves with the operation of the various Danmaku moderation features. Each participant was required to try the basic Danmaku control functions. Because Danmaku is widely built-in feature in most VSPs such as TikTok, Xiaohongshu, Bilibili and WeTV, it was enabled by default to increase participant engagement.

We then provided the five video types to participants, with three videos of each type. Participants were asked to watch at least one video from each genre, with the order of viewing and platforms randomized. While watching each video, we asked participants to freely use the Danmaku moderation feature to adjust the content to their satisfaction for video viewing and explain the reasons for their changes and moderation considerations (e.g., filtering level, etc.). During the continuation of the video watching, we observed how participants set up Danmaku features, how it affected the changes in the Danmaku stream, and how participant adapted or optimized their moderation strategies when they met obstacles. we did not strictly control each participant's video controlling actions (e.g., playback, pause, or playback modes) and Danmaku actions, but rather, wanted participants to watch the video for as long as possible and utilize the moderation functionality.

Participants then utilized the Retrospective Think-Aloud paradigm \cite{van1994think,scheuerman2021framework}. After watching each video, each participant watched a recording of their interaction with the video content and then provided a verbal commentary on their actions and thought processes at the time. The use of retrospective think aloud was chosen as it allowed participants to reflect on their experiences without the pressure of performing the task in real-time, which can sometimes lead to different behaviors or responses. Second, by reviewing the recorded actions, participants could provide insights into their thought processes with the benefit of hindsight, potentially revealing aspects of their decision-making that may not be apparent during concurrent think aloud.

After watching all the videos and thinking aloud, participants were asked about their reasons for not using other moderation features and were asked to think about how to configure and use Danmaku if a limited choice of moderation features existed (e.g., only text filtering or specified content can be used). A debriefing session then allowed me to clarify any ambiguities and provide participants with an opportunity to add any thoughts they might have had difficulty articulating during the task.

\subsection{Data Collection and Analysis}
During this study, audio and video data were recorded from the Retrospective Think-Aloud process when participants were using the technological probes and during the interview questioning. we then transcribed the data for post-experiment analysis. 

To answer the research questions, we used an inductive approach to qualitative thematic analysis \cite{thomas2003general}. Two human-computer interaction researchers with qualitative analysis experience participated in the data analysis as coders. At the beginning of the analysis, the two coders independently read all the data, actively searched for the meaning and pattern of the content, and recorded the analysis memos. Through this scanning process, the coders gradually became familiar with the various types, data volumes, and content distribution of the overall data, and established a preliminary impression of it. Then, the coders discussed their understanding of the data content, the steps of the analysis, and the type of data that needed to be referenced, such as the interview text and the corresponding technical probes, and the design output.

During the process of encoding the data, we used an iterative ``follow-the-leader'' approach, inspired by \cite{agha2023strike}, where we generated an initial topic based on the results of the mutual consultation, and another coder used this initial topic to encode. When there was a question or a new topic in the coding, we were consulted to update the codebook. After many iterations and deliberations, no new topics emerged from any of the data.
Following the interviews, we conducted member-checking by briefly summarizing our understanding of the key points and solicited participant feedback to ensure we had accurately captured the interview content \cite{birt2016member}.

\section{Results from Study 1: Exisitng Limitations and Challenges of Danmaku Moderation on VSP}

The challenges perceived by participants when using the Danmaku moderation features were reflected in participants' subjective experiences, while the limitations of existing Danmaku moderation methods were identified in the challenges they faced at the system and technical levels. Although these perspectives differ, both provided comprehensively insights in Danmaku moderation on VSPs.

\subsection{RQ1: Limitations of Existing Moderation Approaches}
\label{results_study1-1}
Existing moderation methods have significant limitations when dealing with live language and dynamic content, leading to poor participant experiences. These limitations related to transparency, flexibility, the feedback mechanisms that were available, and security and privacy. In addition, there were several moderation-type limitations that existed.

\subsubsection{Lack of Transparency} 

Participants expressed a clear desire for greater insight into how moderation algorithms worked to overcome the  uncertainty and lack of trust they felt. As P1 said, \textit{``It's like a black box. I report something, and I don't know why it was or wasn't taken down. It's hard to trust a system when you can't see how it works.''}. Additionally, participants emphasized the need for clear guidelines and robust feedback mechanisms to foster trust and compliance within the community. When participants had a comprehensive understanding of what behaviors were not tolerated and the reasons for them, they expressed that they would be more inclined to adhere to community guidelines, e.g., \textit{``If I knew the rules clearly, I'd know what to expect. Transparency in moderation would help me understand what's not allowed and why, so I can follow the community guidelines better''} (P7).

\subsubsection{Lack of Flexibility} 

Participants also highlighted the limitations of static moderation rules. A one-size-fits-all approach to moderation was perceived to result in unnecessary censorship or the overlooking of actual violations. This inflexibility could then lead to a mismatch between the moderation action and the context in which the content is presented, potentially affecting user experiences negatively. For example, a comment that might be harmless in one context could be offensive in another, e.g., \textit{``Sometimes a comment might seem fine in one video but not in another. The moderation doesn't seem to get that. It's like it's the same for every video, no matter what the context is.''} (P3).

Additionally, participant frustration stemmed from uniform moderation policies that did not consider the diverse preferences and expectations of the community. As different participant groups may have different tolerance levels for certain types of content, a standardized moderation approach would thus not meet diverse needs. For instance, as explained by P1, \textit{``I feel like the moderation doesn't take into account what different people are okay with. Some people might not mind certain jokes, but others might find them offensive. It's not fair to apply the same rules to everyone''}. 

\subsubsection{Feedback Acknowledgement} 

Participants also reported a lack of system acknowledgment following their reports of inappropriate content. This lack of immediate feedback resulted in a diminished sense of participation and a feeling that their actions were not valued, potentially leading to disengagement from the moderation process. As P3 said \textit{``I've reported stuff before, but I never hear anything back. It's like my report just disappears. A simple 'thank you' or update would make me feel like I'm part of the process.''} Without clear and timely feedback, participants' willingness to moderate thus decreased. This highlights the importance of feedback mechanisms in sustaining participant engagement and reinforcing confidence in the platform's commitment to maintaining a safe and respectful environment. 

\subsubsection{Privacy and Security}

Participants expressed a heightened awareness of the potential risks associated with personal data exposure during the content moderation process. This awareness was noted to potentially deter participants from actively reporting content or participating in moderation activities. As P9 stated \textit{``I'm careful about what I report because I'm not sure who sees my information. If I knew my data was safe, I'd be more willing to help out.''}. Participants also wanted assurance that their privacy was being safeguarded, i.e., \textit{``I want to know what's happening with my reports, but I also don't want to put myself at risk. There needs to be a balance where I feel secure but still informed.''} (P9).

\subsubsection{Moderation Type-Specific Limitations}

\paragraph{Personal Control Moderation.} Participants also reported instances where automated systems failed to accurately analyze the context of content. This inaccuracy led to the inadvertent removal of harmless content and the oversight of objectionable material, which significantly impacted participants' perception of the platform's moderation capabilities. For example, P3 pointed out that \textit{``Sometimes it feels like the system just doesn't get the joke. I've seen comments that were clearly sarcastic get removed, but the real offenders slip through.''}

Participants also noted that the automated systems often misunderstood culturally-specific expressions and language subtleties, leading to inconsistent moderation decisions. For example, P7 said \textit{``We have participants from all over the world, and what's okay to say in one culture might not be in another. The system doesn't seem to take that into account, and it causes a lot of confusion and unfair moderation''}. They thought this inconsistency was particularly evident in content that contained idiomatic expressions, slang, or regional dialects. 

Participants expressed a desire for greater control over moderation settings to customize their content experience. The current systems' limited customization options were identified as a shortcoming to addressing the diverse needs and expectations of participants. The lack of granularity in one's settings was a common point of dissatisfaction. For example, as P6 said \textit{``It feels like we're all treated the same, but not everyone has the same comfort levels. Some might be okay with certain jokes, while others might find them offensive. The settings should let us decide what we want to see and what not.''}. Such settings did not account for the varying tolerance levels and preferences of individuals, e.g., \textit{``I wish there was an option to set my own rules. Like, I don't mind debates, but I don't want to see hate speech. If I could customize the filters, it would make my experience so much better.''} (P5).

\paragraph{Distributed Moderation.}
All participants expressed concerns about the over-moderation of content. They reported that the frequent removal of comments during video viewing disrupted their immersion, causing a shift in focus from the content itself to the moderation process. For instance, P9 noted that \textit{``I was watching a video, and every few minutes there's be a comment disappear. It was like having a hiccup in the middle of a smooth ride. It pulled me out of the moment and took away from the pleasure of just watching.''} This over-moderation was perceived as a hindrance to viewing pleasure, indicating a need for moderation systems to be more nuanced and less intrusive.

\subsection{RQ2: Challenges with Existing Moderation Approaches and Inappropriate Content}
\label{results_study1-2}
This section focus on the challenges participants encountered when using Danmaku moderation methods on VSP. We found participants faced various challenges affecting the effectiveness of comment content management and the participant experience. 

\subsubsection{Maintaining A Seamless Viewer Experience}
There was an ongoing struggle to balance content control with a seamless participant experience, especially with the Distributed Moderation approach. Participants highlighted that moderation systems often failed to strike the right balance, leading to either an overbearing presence that stifled interaction or an insufficient level of control that allowed inappropriate content to persist. For example, P9 said \textit{``It feels like they're (content creators) so worried about anything going wrong that they moderate everything ... but that takes away the natural feel of the conversation. On the other hand, if they don't moderate enough, we see stuff we don't want to. It's a tough balance.''} This comment underscores the importance of developing moderation strategies that respect both the need for content control and the desire for a positive participant experience.

\subsubsection{Harm Caused by the Delay in Inappropriate Material Identification}
Participants also reported that the delay in identifying and addressing inappropriate material can lead to its proliferation, potentially causing harm to the community before any action is taken. P7 believed that the moment one encountered offensive Danmaku comments, harm had already been inflicted: \textit{``When something offensive pops up and stays there, it feels like the platform doesn't care. It's not just about the content, it's about the delay in taking it down.''} There was also a noticeable decline in participants' overall viewing experience due to the cumulative effect of encountering inappropriate content. The repeated exposure to offensive material, which was often not moderated in a timely manner, had led to a heightened sense of frustration and discomfort. This issue not only detracted from the enjoyment of the content but also eroded trust in the platform's commitment to providing a safe and respectful environment, e.g., \textit{``You start to dread what you might see next. It's not just one video; it's every time you come here and get bombarded with stuff you don't want to see, especially, fans content''} (P9).

Furthermore, there was a notable discrepancy between participants' expectations of the moderation process and the actual outcomes they observed. Participants expected rapid and effective moderation, yet they frequently encountered slow and inconsistent responses. P8 mentioned having once muted certain speech, such as match scores, yet similar expressions still prevailed, resulting in a poor viewing experience. Consequently, he felt that the system's reporting and filtering features were utterly ineffective, e.g., \textit{``I expect when I report something, it'll be dealt with quickly. But it feels like my reports are ignored, and the problem persists.''} The gap between expectation and reality can erode trust in the platform and diminish the perceived benefits of participation in moderation efforts.

\subsubsection{Trade-off Between Workload and Benefits}
Participants frequently described participating in content moderation as burdensome. The time and effort required to report and engage in moderation activities was seen as significant, leading to fatigue and a decreased willingness to continue participating. This was especially true then when the perceived impact of their efforts was low.  As P7 said, \textit{``I used to report things all the time, but it started to feel like a second job. It's a lot of work for something that doesn't seem to make a big difference.''}  As P17 said:, \textit{`I ask myself, 'Is it worth it?' Sometimes it feels like my reports are just going into a black hole. I need to see that my efforts are making a difference.''} The burden placed on participants to identify and report content not only affects their viewing experience but also their willingness to contribute to maintaining a healthy community environment. 

\subsubsection{Perceived Censorship}
There were also instances where participants felt their content was misjudged by moderation systems. This led to a sense of injustice and frustration, as participants perceived their legitimate contributions being unfairly censored. Proactive moderation involves taking steps to prevent potential issues based on an analysis of viewer behavior and content, which includes setting guidelines and rules that aim to maintain a healthy community discourse. However, when viewers felt that their legitimate contributions were being unfairly censored, this suggested a disconnect between the proactive moderation strategies in place and the actual viewer experience. Thus, while the moderation system was actively trying to shape the community discourse (proactive stage), it may not be achieving its intended goal of fostering a diverse and vibrant community due to over-zealous or misinterpreted application of its rules. The concerns raised by the participants highlighted the need for a more nuanced and transparent approach to proactive moderation that respects viewer agency and promotes a genuine exchange of ideas. 
As P1 said \textit{``I felt like my voice was being silenced without reason. My comment was taken out of context and deemed inappropriate, even though it was meant to be constructive criticism.''}. Participants also expressed concern over the lack of diversity and vibrant community discourse that could result from such moderation. For example, \textit{``I've seen people self-censor because they're unsure if their comment will pass moderation. It's a shame because it stifles the variety of opinions we could have in the community''} (P6).

\subsubsection{Moderation Detracting from One's Viewing Experience}

Participants reported that moderation actions, particularly when they were frequent or visibly intrusive, could significantly disrupt the viewing experience. The sudden removal of comments or constant notifications about moderated content detracted from the video content itself. For instance, P4 stated \textit{``I was really into the video, and then suddenly a bunch of comments disappeared. It was jarring and took me out of the moment. It felt like the platform was more focused on moderation than my viewing experience.''}. As P3 said, \textit{``I understand the need to keep the content clean, but sometimes it feels like the moderation is too heavy-handed. It takes away from the natural flow of the conversation and the enjoyment of the video.''}

\subsection{Summary}
This section contributed a comprehensive examination of the moderation challenges and limitations associated with Danmaku systems on video-sharing platforms. Our findings highlighted the inherent limitations of existing moderation approaches, such as a lack of transparency, rigid rules, and inadequate feedback mechanisms, which undermine user trust and engagement. Our research underscores the need for moderation strategies that are sensitive to the dynamic,  contextual nature of Danmaku. By revealing the gap between current practices and user expectations, this study lays the groundwork for the development of more effective and user-centered moderation solutions, which are explored in RQ3. It also emphasizes the importance of incorporating user perspectives and cultural considerations into the design of moderation systems, paving the way for a more inclusive and responsive approach to content governance in digital spaces.

\section{Study 2: Co-design Workshop}
\label{CodesignWorkshop}
Following the insights gained from \autoref{ProbInterview}, we conducted co-design workshops to understand \textit{``What design opportunities do participants identify to enhance Danmaku moderation methods?''}. The co-design workshops were crucial for translating the identified challenges from \autoref{ProbInterview} into actionable design solutions. These workshops utilized the same participants as in \autoref{ProbInterview}. During these workshops, we provided materials to support their understanding and exploration of potential Danmaku moderation methods for video sharing platforms (VSPs). This user-centered approach was chosen to increase the likelihood that the resulting solutions would align with actual user needs rather than relying on assumptions.

As noted, the same people that participated in the study in \autoref{ProbInterview} also participated in the co-design workshops described herein. By first exploring user experiences and challenges through the interviews in \autoref{ProbInterview}, and then engaging the same participants in co-design activities in this study, we were able to undertake a more nuanced and user-validated exploration of potential design opportunities for Danmaku moderation methods that would meet user needs and expectations.

\subsection{Procedure and Activities}
Participants were divided into groups of three. To avoiding any bias associated with familiarity and similarity of backgrounds, the assignment was randomized and tried to ensure that the participants in each group did not know each other. Inspired by prior co-design studies \cite{agha2023strike,yen2023storychat}, three activities were used: 1) creating storyboards of the existing limitations and challenges identified in \autoref{ProbInterview}, 2) group discussions based on end user needs and potential solutions, and 3) developing a low-fidelity prototype of a design-guided feature for Danmaku moderation. We provided samples, tips, and guidelines for these design activities, following existing training methods for storyboard creation\footnote{https://www.uxdesigninstitute.com/blog/ux-storyboard/}, visualizing user experience ideas\footnote{https://www.nngroup.com/articles/storyboards-visualize-ideas/} and prototyping\footnote{https://www.interaction-design.org/literature/topics/prototyping}. Considering the diverse backgrounds and different levels of design skills of the participants, we prepared materials that supported low to medium-high fidelity prototyping (e.g., paper and colored pencils, computer slides or online drawing tools, interactive design tools Figma or Sketch or Adobe XD, and so on). The goal was to encourage participants to use tools they were comfortable with to fully express their personal design concepts, rather than being limited by them.

\subsubsection{Activity 1: Storyboard Creation}
At the beginning of this activity, storyboard templates were used to guide participants through the ideation and design process of Danmaku moderation solutions. These templates were crafted to incorporate a range of visual and textual elements that would assist participants in effectively articulating their conceptualizations and design processes. The storyboard templates provided a structured yet flexible framework, enabling participants to sketch out their ideas, integrate relevant screenshots or interface elements, and annotate their thought processes and intended participant interactions. We included sections for participants to define the problem statement, outline participant scenarios, and detail the proposed moderation features, ensuring that each design solution was grounded in participant-centered considerations. 

Each participant was asked to create their own storyboard based on scenarios of Danmaku interactions that were identified in \autoref{ProbInterview}. At this stage, participants were provided with paper storyboard templates, visualizations of participants' outward behaviors and references to their internal reactions to Danmaku. Throughout the activity, we guided participants, using the storyboard as a blueprint for the narrative, to help organize and articulate key visual elements. At the same time, we asked participants relevant questions about their design motivations to understand their scenarios and proposed solutions. 

\subsubsection{Activity 2: Group Discussion}
During this activity, each participant was asked to present their storyboard to their group. They were required to articulate the motivation for their design, its need, and the potential reaction to the Danmaku. Other participants and the researcher provided feedback on each storyboard, including its strengths and weaknesses, potential perceptions of viewers, and potential solutions.

\subsubsection{Activity 3: Prototyping Danmaku Moderation Features}
Based on the content of the storyboards and the results of the group discussions, during this activity, participants were first asked to create prototypes using design tools to conceptualize the details of their proposed solution for Danmaku moderation. During this time, we assisted the participants in using the tools, responded to requests for help, and helped them to organize and structure their ideas. Each participant was given half an hour to prototype their design. After completing their prototype, participants demonstrated it to the group by explaining the Danmaku that needed to be reviewed, discussing their solution, and indicating how this solution would impact viewers. After sharing, group members asked questions and provided relevant feedback on the design.

\subsection{Data Collection and Analysis}
During this study 27 design ideas were collected (\ref{tab:design_examples}), in addition to audio and video recordings. We transcribed all the audio data. To identify interview topics, we used an inductive approach to qualitative thematic analysis \cite{thomas2003general} similar to \autoref{ProbInterview}. To answer RQ3, we analyzed the prototyped features and conceptual interactions detailed in the co-design artifacts and used relevant data from \autoref{ProbInterview} as secondary materials for our analysis. Finally, we used the card sorting method \cite{aarts2020design} to classify the design output of the challenges and limitations found in \autoref{ProbInterview} to determine design opportunities.

\begin{table}[!ht]
    \centering
    \resizebox{\textwidth}{!}{
    \renewcommand\arraystretch{2.5}
    \begin{tabular}{cll}
    \toprule[1.5pt]
     \textbf{Example} & \textbf{Design Solutions}  & \textbf{Description}\\ 
    \toprule[1.5pt]
     1 &  \makecell[l]{Scaffolding Supports \\for Pre-setting Moderation Rules}  
       &  \makecell[l]{Implement a user-friendly interface that allows users to pre-set moderation rules \\according to their preferences or the content requirements of the video. \\This could include setting the strictness level, filtering out specific types of content, or defining community guidelines.}\\ \hline
     2 &  Natural Language-Based Interactions   
       & \makecell[l]{Develop a system where users can interact with the moderation tools using natural language processing (NLP). \\This would make the moderation process more intuitive and conversational, similar to chatting with an AI like ChatGPT.}\\ \hline
     3 &  Collaboration with AI Agents  
       &  \makecell[l]{Integrate AI agents to assist users in the moderation process, \\helping to scale the efforts and manage a larger volume of content. \\AI agents can learn from user interactions and improve the moderation over time.}\\ \hline
     4 &   Simplicity in Moderation Interface 
       & \makecell[l]{Design and optimize the moderation features to be as simple as a chat interface,\\making it easy for users to engage with the tools without a steep learning curve.}\\ \hline
     5 & Visualization of Content 
       & \makecell[l]{Use visual representations to differentiate between negative and positive content, displayed alongside the video.\\This can help users quickly assess the sentiment of the Danmaku.}\\ \hline
     6 & Algorithm Precision Improvement   
       & \makecell[l]{Enhance the algorithm to accurately identify inappropriate information by considering the video content \\and current internet trends, reducing the misjudgment of context.}\\ \hline
     7 &  User Interaction with Content  
       & \makecell[l]{Allow users to like or dislike Danmaku content, with liked Danmaku being highlighted. \\This promotes positive interaction and gives users more control over the content they see.}\\ \hline
     8 &  Cultural Sensitivity and Joke Moderation  
       & \makecell[l]{Provide platform responses and prompts for inappropriate jokes or cultural references, \\emphasizing the importance of Danmaku etiquette.}\\ \hline
     9 & \makecell[l]{Automatic Extraction \\and Explanation of Internet Slang}   
       & \makecell[l]{Automatically identify and explain internet slang or memes within \\the Danmaku to enhance viewer understanding and participation.}\\ \hline
     10 &  Personalized Danmaku Content  
        & \makecell[l]{Let users pre-set common Danmaku phrases for quick sending, and set up keyword blocking to filter out unwanted content.}\\ \hline
     11 &  Real-Time Dislike and Blocking  
        & \makecell[l]{Enable users to block danmaku with a simple dislike gesture or by selecting specific words, \\enhancing the control over their viewing experience.}\\ \hline
     12 &  Gesture and Expression Recognition  
        & \makecell[l]{Utilize gesture or facial expression recognition to dynamically adjust the moderation settings \\based on the user's reactions to the content.}\\ \hline
     13 &  Anonymity and Protection Mechanisms  
        & \makecell[l]{Clearly display and explain the anonymity protection mechanisms in place to reassure users of \\their privacy during the moderation process.}\\ \hline
     14 &  Flexible Moderation for Cultural Nuances  
        &  \makecell[l]{Adapt moderation strategies to accommodate different cultural backgrounds \\and user-generated content nuances, such as ethnic sensitivities or jokes.}\\ \hline
     15 &  Community Contribution and Rewards  
        & \makecell[l]{Highlight user contributions to the Danmaku community and \\provide incentives to foster a sense of belonging and engagement.\\Recognize user efforts with rewards when appropriate.}\\ \hline
     16 &  Personalized Moderation Settings  
        & \makecell[l]{Allow users to decide and control the content they are exposed to \\through personalized moderation settings, enhancing user autonomy.}\\ \hline
     17 &  Intelligent Learning from User Interactions  
        & \makecell[l]{Enable the system to learn from user interactions with Danmaku \\and adjust moderation accordingly, catering to individual sensitivities and priorities.}\\ \hline
     18 &  Unobtrusive Moderation During Viewing  
        & \makecell[l]{Avoid any mandatory moderation displays during video viewing to maintain the integrity of the viewing experience.}\\ \hline
     19 &  Increased Visibility of Impact 
        & \makecell[l]{Enhance the visibility of the impact of user moderation efforts, \\such as displaying the number of reports and praises, to motivate users.}\\ \hline
     20 &  Context-Aware Moderation  
        & \makecell[l]{Enable the system to understand the nuances of different content environments and adjust moderation settings accordingly,\\such as differentiating between humorous and serious videos.}\\ \hline
     21 &  Vague Directives for Moderation  
        & \makecell[l]{Allow users to give vague directives that the system can \\interpret to moderate Danmaku, reducing the need for fine-tuning each moderation feature.}\\ \hline
     22 &  Simplified Reporting Process 
        & \makecell[l]{Streamline the reporting process using AI to categorize \\and prioritize reports, ensuring faster responses and easier management.}\\ \hline
     23 &  Community-Specific Moderation Preferences  
        & \makecell[l]{Foster community-driven moderation preferences, \\especially in videos with a strong fan base, catering to the specific Danmaku moderation tastes of the community.}\\ \hline
     24 &  Feedback on Reporting Actions  
        & \makecell[l]{Provide feedback on user reporting actions, including regular reports on \\the handling of reports and clear consequences for those reported.}\\ \hline
     25 &  Transparency in AI Filtering Mechanisms  
        & \makecell[l]{Clearly explain the AI-related filtering mechanisms, with examples, \\to help users understand different intelligent filtering levels.}\\ \hline
     26 &  Reducing Misjudgments  
        & \makecell[l]{Improve system intelligence to minimize misjudgments, ensuring that benign content or internet memes \\that contribute to the atmosphere or user engagement are not mistakenly removed.}\\ \hline
     27 &  AI Companionship  
        & \makecell[l]{Introduce an AI companion that empathizes with viewers, responding to Danmaku content to mitigate discomfort \\from negative Danmaku and enhance engagement with interesting ones.}\\ \hline

    \end{tabular}
    }
    \setlength{\belowcaptionskip}{3pt}
    \caption{Design Solutions in Co-design workshop}
    \label{tab:design_examples}
\end{table}

We analyzed the design artifacts based on four types of common content moderation approaches (e.g., active, proactive, reactive or passive \cite{habib2019act}) and categorized them based on the moderation concepts, inspired by card sorting \cite{lee2019design}. Proactive moderation is an anticipatory approach, where moderators predict potential issues based on an analysis of participant behavior and content on the platform and then take preemptive measures (e.g., using machine learning models to identify and intercept potentially harmful content or intensifying moderation efforts before specific events or holidays). Reactive moderation involves moderators intervening only after participants report issues or the system automatically detects potential problems. This moderation type relies on participant feedback and alerts from automated tools, with moderators responding to, and resolving, issues rather than preventing them. Active moderation refers to the proactive search and assessment of content on a platform by moderators to ensure it adheres to community guidelines and policies. Here, moderators not only respond to participant reports or system prompts but actively browse content to identify potential issues. Passive moderation typically implies a less vigorous moderation process, possibly due to resource constraints, insufficient technology, or other reasons that prevent moderators from actively or promptly addressing content issues. With passive moderation, harmful content may spread more widely before being identified and addressed.

\begin{figure}
    \centering
    \Description{A figure showing four design artifacts mapped onto a two-dimensional framework. The framework dimensions are proactive vs reactive moderation on one axis and high vs low participant engagement on the other. The examples include: (10) a proactive moderation solution with high participant engagement, (9) a proactive moderation solution with low participant engagement, (3) a reactive moderation solution with high participant engagement, and (19) a reactive moderation solution with low participant engagement.}
    \includegraphics[width=1\linewidth]{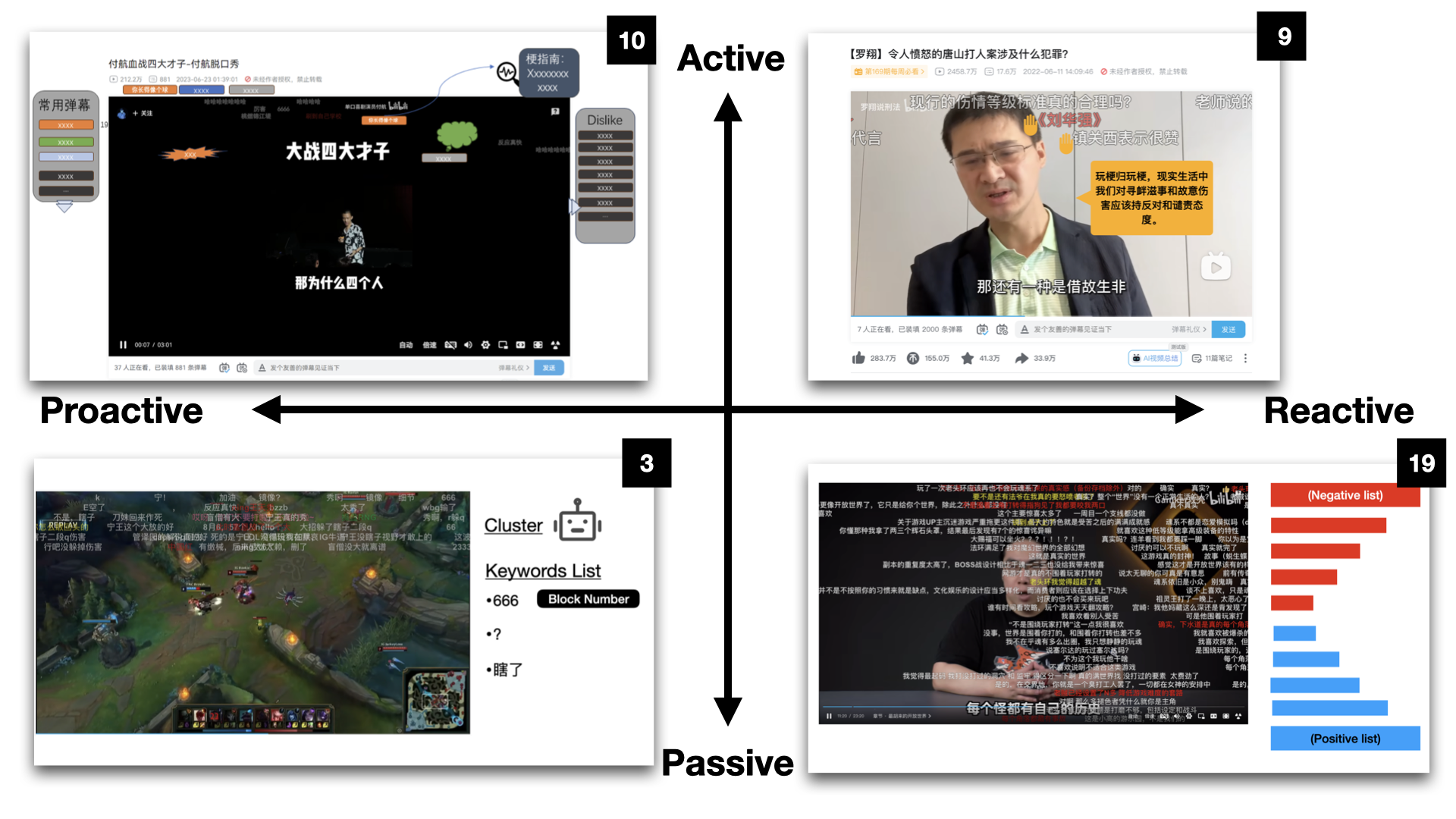}
    \caption{Examples of four design artifacts mapped to the dimensions of the framework. For example, proactive moderation with high participant engagement (10), proactive moderation with low participant engagement (9), reactive moderation with high participant engagement (3), and reactive moderation with low participant engagement (19).}
    \label{fig:design_examples}
\end{figure}

Using these four types of content moderation, two researchers then created an analytical framework consisting of the required level of participant engagement (i.e., passive and active) and the timing of actions (i.e., proactive and reactive actions). The researchers then mapped the design artifacts to this framework (\autoref{fig:design_examples}) and categorized them based on their intervention type: preventive, precautionary, participatory, or administrative. This process involved an in-depth analysis of each intervention's function, participant interaction needs, and its role in preventing or responding to Danmaku moderation. The researchers analyzed each moderation method for its strengths and potential weaknesses, taking into account their feasibility and user acceptance in practice. These results were then integrated into a strategy framework that encompassed the different types of Danmaku moderation to provide clear guidelines for designers and content moderation researchers.

\section{Results from Study 2: Design Opportunities of Danmaku Moderation}
\label{results_study2}
The findings are organized into two sections, i.e., design opportunities for Danmaku moderation on VSPs based on the analytical framework, and design guidelines. 

\subsection{Proactive Moderation with Active Engagement}
When participants are actively engaged and wish to proactively take initiative in content moderation, they desire more control over the moderation process. They suggested features that went beyond traditional mechanisms, introducing novel concepts to enhance user participation and content quality. While participants still valued traditional features like liking or disliking comments and gamification rewards, they envisioned these being integrated into more sophisticated systems. They thus, \textit{``want to like or dislike comments to highlight the best or the worst ones''} (P1). Therefore, highly liked comments should be being highlighted to foster active participation. Such a feature could help surface the most relevant and positive comments, enhancing community engagement and content quality.

Participants also suggested a community co-created content filter, a collaborative tool allowing user groups to collectively define and adjust content filtering rules. P12 elaborated, \textit{``Imagine if we could work together to create smart filters that learn from our community's norms and evolving language patterns.''} This suggestion transcends simple keyword filtering by leveraging machine learning algorithms to identify and adapt to changing linguistic patterns and community standards.

Simplifying the moderation process was another key theme. Participants wanted to provide vague or general directives for moderation, which a system could interpret and execute without requiring detailed adjustments. For example, as P8 said, \textit{``I should be able to give general instructions like `block negative comments' ... instead of adjusting precise requirements step by step across different buttons.''}. This suggestion points towards the need for a more user-friendly interface powered that can interpret and execute vague directives without requiring detailed adjustments. Simplifying the moderation process could empower more participants to actively participate without needing in-depth knowledge of the system. 

Recognition and rewards were also found to be strong motivators. For example, \textit{``seeing my contributions and earning badges would make me want to moderate more''} (P16). This indicates a desire for a more sophisticated reward system that reflects genuine contributions to community well-being. Systems should thus display participant contributions, such as the number of moderations and successful reports, and reward them using points or badges. Such systems could enhance participant motivation and satisfaction, leading to more active and consistent participation in moderation activities.

Improving the ease of reporting was also a common idea, but participants envisioned more than just streamlined forms. As P6 explained \textit{``Reporting should be quick and easy; I don’t want to fill out long forms''}. Systems should thus streamline the reporting process to automatically classify reported content, requiring minimal input from participants and ensuring swift responses. Simplifying reporting can encourage more participants to report inappropriate content, leading to a safer and more enjoyable viewing environment.

\subsection{Proactive Moderation with Passive Engagement}
There were also several passive engagement, proactive moderation ideas as well. P17 mentioned that because dialogue is one of the most natural and convenient methods of moderation interaction, \textit{`I want to give simple commands like I would in a chat with ChatGPT''}. Simplifying moderation functions to resemble interactions with LLMs, where participants could input straightforward commands like "block all offensive comments", would lower the barrier for participants who might be less engaged or tech-savvy.

As most participants appreciated contextual help (e.g., \textit{``Explain the trending slang in comments so I can understand and join in''} (P9)), they wanted a VSP to automatically identify and extract trending internet slang from Danmaku comments and display explanations alongside the comments to help them understand and participate in discussions.

Techniques to limit disruptions were also mentioned. As P1 said,  \textit{``Don’t disrupt my viewing experience with moderation prompts''}. Participants wanted moderation prompts to be optional pop-ups or displayed after a video ended to ensure that the viewing experience would remain immersive and uninterrupted.

Some participants were intrigued by the idea of non-verbal cues. For example, as P14 said \textit{``it would be cool if the system could block comments based on my reactions''}. She wanted gesture recognition, facial expression detection, or eye-tracking technology to automatically detect and block Danmaku that caused her discomfort. Such methods could provide a seamless and non-intrusive way to filter out unwanted content, enhancing participant comfort and satisfaction.

\subsection{Reactive Moderation with Active Engagement}
With an active level of engagement but reactive moderation, it is key that moderation does not interfere with user viewing experiences. To overcoming the complexity of manual moderation settings and reducing the learning curve for new users, many participants expressed a desire for intuitive and participant-friendly interfaces that would enable them to preset moderation rules (e.g., \textit{``I want templates that I can just pick and slightly adjust to fit my preferences'} (P10)) or comments (e.g., \textit{``having quick access to my frequently used comments would be fantastic ''} (P13)). 

Since AI or LLM tools can handle large volumes of comments, they were noted as being able to potentially improve the efficiency of content moderation (e.g., \textit{``If an AI can do the initial screening, I can just review what needs my attention''} (P17)). The development of advanced algorithms capable of accurately understanding both video content and the latest internet trends were also noted as being potentially useful (e.g.,\textit{``The system needs to understand context better to avoid deleting good comments''} (P18)). 

\subsection{Reactive Moderation with Passive Engagement}
Viewers who are passively engaged and want to reactively moderate content require less control over the moderation process. For these viewers, it was recommended that natural language processing should be used to enable them to express their moderation intentions simply, with a system automatically adjusting settings for them. Ease of use was also a recurring theme (e.g., \textit{``I want to just tell the system what to do in plain language''} (P2)) because such users were perceived as being not tech-savvy and thus required a more assistive approach to moderation. 

Participants also wanted to see the results of their actions. As P7 said, \textit{``I wish it would show how my reports help can motivate me to report more''}. Knowing how their actions contribute to a safer and more positive community could be a strong motivator for some users, as could displaying feedback about the impact of participant moderation actions, such as the number of reports or praises received. 

Reducing false positives was also a priority (e.g., \textit{``The system should be smart enough to understand the context and not delete positive comments''} (P20)). Developing smarter systems that can accurately interpret the context and semantics of Danmaku comments to avoid misjudgments would reduced the risk of valuable comments being incorrectly flagged or removed, thus improving the overall user experience and trust in moderation systems.

\subsection{Design Guidelines}
\label{design_guidelines}

\subsubsection{Participatory (proactive with high engagement)}
The guidelines should encourage viewers to actively participate in the management of Danmaku content and work together to create a healthy environment for communication.

\begin{itemize}
    \item Create incentives to reward viewers who actively report inappropriate content and contribute positive comments \cite{hron2022modeling}.
    \item Develop interactive features, such as ``Like'' or ``Step on'', to allow viewers to rate Danmaku content, so that good content gets more opportunities to be displayed.
    \item Set up community forums or workshops to encourage viewers to participate in the discussion and formulation of Danmaku management policies, so as to enhance viewers' recognition of the platform's rules \cite{seering2019moderator}.
    \item Provide a transparent moderation feedback mechanism, allowing viewers to understand the progress and results of the handling of reports and ensuring the transparency of the moderation process.
\end{itemize}

\subsubsection{Administrative (reactive with high engagement)}
Guidelines for administrative interventions should focus on post-post Danmaku content management and response mechanisms.

\begin{itemize}
    \item Establish clear community guidelines and publicize detailed categorization of offending content and corresponding penalties \cite{seering2020reconsidering}.
    \item Establish an efficient user reporting system that enables viewers to quickly report inappropriate content \cite{gorwa2020algorithmic}.
    \item Provide comprehensive training for administrators to ensure that they are able to handle offending content fairly and effectively \cite{straub1998coping}.
\end{itemize}

\subsubsection{Precautionary (proactive with low engagement)}
In terms of early warning interventions, design guidelines should focus on preventing potentially inappropriate Danmaku content in advance.

\begin{itemize}
    \item Implement user authentication mechanisms to ensure that there is a traceable individual behind each account \cite{king2012modifying}.
    \item Design real-time monitoring algorithms capable of recognizing potentially offending content before Danmaku is posted \cite{wang2022ml,gorwa2020algorithmic}.
    \item Create a content flagging system that automatically flags suspicious content for subsequent manual moderation \cite{link2016human,hron2022modeling}.
\end{itemize}

\subsubsection{Preventive (reactive with low engagement)}

Preventive interventions should focus on reducing the generation and distribution of inappropriate Danmaku content.

\begin{itemize}
    \item Provide tools that allow viewers to customize filtering rules, such as blocking specific keywords or viewers \cite{jhaver2023personalizing}.
    \item Educate viewers about netiquette and guidelines for using Video Sharing Platforms to increase their awareness of self-moderation \cite{seering2020reconsidering}.
    \item Educate viewers about what content may be deemed inappropriate through alert messages before they post Danmaku.
\end{itemize}

\subsection{Summary}
This section illuminates how one-size-fits-all moderation approaches fundamentally fail to adequately address the contextually embedded nature of Danmaku on VSPs. It emphasized the importance of moderation systems and clear definitions of objectionable content that can adapt to evolving cultural norms and different types of content. The insights reveal strategies for developing advanced AI capabilities with nuanced context awareness for real-time user multimedia. By centering end-user voices through inclusive participatory processes, this co-design activism exemplifies how human-centered approaches can transform digital governance frameworks. This transformation aims to cultivate vibrant, safe online spaces that respect diverse socio-cultural contexts, enable self-expression, and mitigate toxic interactions.

\section{Discussion}
In this section, we reflect on the specific findings of our study within the context of Danmaku content moderation, contrasting these with existing literature to elucidate both commonalities and distinctive aspects. 
Our study provides a comprehensive examination of the unique challenges and opportunities within the Danmaku content moderation landscape. While there are overlaps with general content moderation issues, the specific attributes of Danmaku demand a distinct set of strategies that are both technically adept and culturally aware, fostering a more inclusive and interactive user environment.

\subsection{Rethinking about RQ1.Limitation and RQ2.Challenges}
This research examined the unique challenges and opportunities that exist when designing systems to assist with the moderation of Danmaku. While there are overlaps with general content moderation, Danmaku demands a distinct set of strategies that are technically adept and culturally aware, to foster a more inclusive and interactive environment. We shedded light on the challenges in moderating Danmaku (Section\ref{results_study1-1}), which stem from the real-time and dynamic nature of the content. Danmaku's dynamic nature necessitates the need for a more context-sensitive, agile moderation approach compared to other social media or video platforms. Our findings emphasized the importance of developing moderation systems that can analyze both text and video content, as well as user reactions, in real time. Addressing this requirement will be crucial for improving moderation. Moreover, our research expands on previous research by underlining the significance of prompt responses in live streaming, as noted in \cite{cai2021moderation}.

The perceived challenges in Section\ref{results_study1-2} highlight how participant dissatisfaction stemmed from the inflexibility of moderation rules and their impact on the viewing experience. Our study participants expressed a desire for more personalized moderation strategies that respected the unique culture of Danmaku while ensuring a safe environment. This resonates with the broader discourse on participant-generated content moderation \cite{jhaver2019human, jiang2023trade}, but our findings were particularly tailored to the anonymous and community-driven nature of Danmaku interactions.

Danmaku presents a unique form of participant-generated content on VSP, distinguishing itself from traditional social media posts or video comments. The immediate, fluid nature of Danmaku creates a dynamic viewing experience that sets it apart from other online contexts. Unlike static comments or posts studied by Cai et al. \cite{cai2021moderation} in live streaming platforms or Jiang et al. \cite{jiang2023trade} in video comment sections, Danmaku comments are intimately connected to specific moments in the video content, creating a real-time, interactive social dimension \cite{wu2019danmaku,ma2017video}. While Cai et al. \cite{cai2021moderation} emphasized the importance of immediate response times in live streaming and Jiang et al. \cite{jiang2023trade} demonstrated the benefits of personalized moderation strategies, Danmaku moderation requires an even more nuanced understanding of the content's relationship to the video timeline. Moderation systems must not only analyze the textual content but also comprehend the video content and viewers' immediate reactions to achieve accurate and efficient content management \cite{cai2021moderation,cai2022understanding}.

The global nature of video platforms introduces additional complexity to Danmaku moderation. As noted by Yang et al. \cite{yang2021danmaku}, Danmaku often encompasses a wide range of languages and cultural backgrounds, reflecting the diversity of its global user base. This diversity presents a significant challenge in maintaining content inclusivity while preventing the dissemination of harmful or offensive material \cite{yang2020danmaku}. Furthermore, effective Danmaku moderation must balance user privacy and freedom of expression with content control, a crucial factor in maintaining user trust and platform reputation. This transforms Danmaku moderation from a purely technical challenge into a complex issue involving social, cultural, and ethical considerations.

\subsection{Rethinking about RQ3. Design Opportunities}
In the \autoref{CodesignWorkshop}, we identified the need for participant-centric moderation tools that enhance community engagement, such as likes or dislikes, and reward mechanisms that acknowledge participant contributions to the moderation process. Furthermore, our findings stressed the importance of cultural sensitivity in moderation practices, aligning with the global and diverse participant base of video-sharing platforms. These opportunities build upon existing knowledge on community-based moderation \cite{seering2020reconsidering, seering2019beyond}, introducing novel strategies that are uniquely suited to the Danmaku context.

Base on previously studies about content moderation challenges and limitations in \autoref{ProbInterview}, these workshops revealed a spectrum of user perspectives regarding the allocation of Danmaku moderation responsibilities, which significantly influenced user behavior and engagement with moderation processes. These diverse viewpoints underscored the need for a clear delineation of responsibilities among platforms, users, and potential third-party moderators, aligning with findings from Bamberger et al. \cite{bamberger2021allocating}. A key finding was the varying levels of user involvement in moderation activities. Some participants expressed a strong desire for active engagement in the moderation process, viewing it as a community responsibility. For instance, P7 stated, \textit{``I feel it's our duty as users to report inappropriate content and help maintain a healthy environment''}. These participants were more likely to utilize reporting tools and participate in community-driven moderation initiatives, echoing the importance of user participation highlighted by Gorwa et al. \cite{gorwa2020algorithmic}.

Conversely, another group of participants preferred a more hands-off approach, expecting platforms to handle the bulk of moderation tasks. As noted by P12, \textit{``I don't want to spend my time moderating content. That's what the platform should be doing.''} These participants were less likely to engage with moderation tools or report problematic content, relying instead on automated systems and platform-led initiatives, similar to the expectations discussed in Seering et al.'s work \cite{seering2020reconsidering}. These workshops also uncovered a nuanced middle ground, where participants supported a collaborative approach to moderation. P3 suggested \textit{``I think it should be a joint effort. The platform provides the tools, and we users help identify issues they might miss''}. This group showed a willingness to participate in moderation but also expected significant support and clear guidelines from the platform, aligning with the balanced approach advocated by Cai and Zhao \cite{cai2021moderation}. These varying perspectives directly impacted participant behavior. Participants who viewed moderation as a shared responsibility were more likely to actively report inappropriate content, engage with customizable filters, and participate in community discussions about content standards. In contrast, those who placed the onus primarily on platforms showed lower engagement with these tools and were less likely to report violations unless they were particularly egregious.

Interestingly, cultural backgrounds played a role in shaping these perspectives. Participants from collectivist cultures, like P9, tended to favor more centralized, platform-led moderation approaches: \textit{``In our culture, we expect authorities to maintain order. The same should apply online''}. On the other hand, those from more individualistic backgrounds, such as P15, often advocated for decentralized, user-driven moderation: \textit{``I believe in the wisdom of the crowd. Users should have more say in what's acceptable.''} This aligns with the theoretical framework proposed by Fesler \cite{fesler1965approaches} on the influence of cultural perspectives on governance approaches. These findings highlight the need for platforms to develop flexible moderation strategies that can accommodate diverse user expectations and cultural backgrounds. For instance, offering both automated moderation tools and user-driven reporting systems could cater to different user preferences. Additionally, platforms should consider providing cultural sensitivity training for moderation teams to better navigate the complexities of global content, as suggested by Yang et al. \cite{yang2021danmaku}.

These varying perspectives also underscore the importance of clear communication from platforms about moderation policies and user roles during moderation. Educating users about their potential impact on content moderation could encourage more active participation from those who might otherwise remain passive, a point emphasized in the work of Jiang et al. \cite{jiang2023trade}.
    
\subsection{Allocation of Responsibilities for Danmaku Moderation}
Our experiments have shed light on the diverse perspectives individuals hold regarding the moderation responsibilities of Danmaku content. These views significantly influence viewers' behavior and their expectations of the moderation function. This finding aligns with existing work on the allocation of responsibilities for content moderation \cite{bamberger2021allocating}. It underscores the necessity for a clear exploration and delineation of the responsibilities of platforms, users, and third-party moderators.
In terms of Platform Responsibilities, these platforms have a foundational role in Danmaku content moderation. They must provide effective automated and manual moderation tools to ensure non-compliant content is addressed promptly and moderation algorithms and policies are updated regularly. For instance, studies have highlighted the importance of platforms' proactive stance in developing tools that can preemptively identify and filter inappropriate content \cite{gorwa2020algorithmic}. Additionally, platforms should encourage user participation in moderation through reporting systems and by enhancing users' self-management capabilities, such as customizable Danmaku filters.
On the other hand, users are not merely consumers of content but active participants in the moderation process. They are encouraged to contribute to content moderation by flagging inappropriate content, which is vital for platforms to maintain a healthy community environment. User education on netiquette and guidelines for responsible behavior is crucial to foster a culture of self-moderation and community stewardship.
The involvement of third-party moderators can complement the efforts of platforms and users. These entities can offer specialized services to assist with the moderation process, bringing in additional expertise and resources. However, their role also raises questions about accountability and transparency in decision-making processes.
In terms of the legal and ethical responsibilities, Content moderation encompasses not only legal obligations but also ethical considerations. Platforms must navigate the complex landscape of content regulation, balancing the need to prevent the dissemination of harmful content while upholding principles of free speech and privacy. The ethical implications of automated versus manual moderation have been discussed in the literature, with a call for a balanced approach that respects user agency and cultural diversity \cite{seering2020reconsidering}. If applicable, moderation should clarify the legal responsibilities of social media platforms in terms of content regulation and ensure that platforms take appropriate measures to respond to inappropriate content.

\subsection{Limitations and Future Work}
The drawback of this study is that due to the limited experimental site, most of the participants are local users, whose preferences and habits on Danmaku content moderation may be influenced by region, culture, and language. However, our work as the first to explore the moderation of Danmaku content is also beneficial to serve as an inspiration for other regional studies. In future research, we may explore ways to incorporate the moderator perspective and will try to further explore the impact of different types of design guidelines on moderation of Danmaku content and system development by looking at different types of design guidelines. 
Clarifying and distributing responsibilities among different stakeholders are essential for effective Danmaku content moderation. It allows for the protection of free expression while ensuring a safe and engaging user environment. We would continue to explore the nuances of responsibility allocation across different cultural and legal contexts, contributing to the development of inclusive and adaptive content moderation frameworks in the future work.

\section{Conclusion}
Our study provides insights into the user experience and content moderation of Danmaku on video sharing platforms. Our findings reveal the limitations and challenges of current Danmaku moderation methods, especially when dealing with high-speed, dynamic Danmaku content. User dissatisfaction with moderation rules and the impact on the video experience suggest important design guidelines. Our research emphasizes the need for more flexible, user-driven moderation methods that accommodate the unique culture of Danmaku while safeguarding a safe, interactive user environment. Our results not only expand the understanding of Danmaku content moderation, but also provide valuable insights into the design of more effective Danmaku management on video sharing platforms.


\bibliographystyle{ACM-Reference-Format}
\bibliography{references}

\end{document}